\documentclass[runningheads]{llncs}
\usepackage[T1]{fontenc}
\usepackage{graphicx,amsmath,amsfonts,amssymb,amsthm,geometry,afterpage} 
\usepackage{float,dblfloatfix}
\usepackage{authblk}
\usepackage{booktabs,tabularx,adjustbox,rotating}
\usepackage{algorithm}
\usepackage{algpseudocode}
\usepackage{longtable}
\usepackage{xcolor}   
\usepackage{amssymb}
\usepackage{graphicx} 

\begin{document}
\newcolumntype{C}{>{\centering\arraybackslash}p{1em}}
\newcolumntype{D}{>{\centering\arraybackslash}p{1.3em}}
\title{3 Players Auction Bridge - Statistical Algorithmic Strategies}
%
%
\author{Sourish Sarkar\inst{1}\orcidID{0009-0002-1887-0007} \and
Aritrabha Majumdar\inst{2}\orcidID{0009-0003-9149-0729} \and
Moutushi Chatterjee\inst{3}\orcidID{0000-0003-4848-1627}}
\authorrunning{S.Sarkar et al.}
%
\institute{Indian Statistical Institute, Hyderabad \email{sourish.sarkar13@gmail.com}\and
Indian Statistical Institute, Bangalore
\email{bmat2311@isibang.ac.in} \and
Indian Statistical Institute, Bangalore
\email{tushi.stats@gmail.com}}
\maketitle              
\begin{abstract}
   Three-player auction bridge is an updated version of the worldwide-recognized auction bridge card game. Dynamic partners and opponents are the key factors in this game. This dynamic property reduces the bias and increases the authenticity of this game. The article provides insight on statistical learning and insight for a three-player auction bridge game. An exact algorithm for various scenarios is developed along with the corresponding winning strategies to explore different game plannings and the computational intelligence involved. Based on the algorithm, we simulate data and define the probabilistic values to win a certain game. 
\keywords{Auction Bridge\and Bidding Strategy\and Computational Intelligence\and Multiplayer Game\and Planning\and  Statistical Analysis and Learning}
\end{abstract}
\section{Introduction}

The three-player auction bridge is an advanced version of the traditional auction bridge. In this game, the bidding section point table differs from that of the conventional auction bridge, as discussed in Sarkar et al.~\cite{ourpaper}. The unique aspect of this variant is that opponents and partners are dynamic, ensuring that there is no bias.

Whitehead~\cite{whitehead1930auction} was among the first to document principles and game rules for auction bridge. Delooze and Downey~\cite{inproceedings} discussed the complexity of the auction bridge game as a multi-player game and proposed a neural network-based self-organizing map for bidding the so-called 'No Trump hand' effectively. Chen et al.~\cite{10960098} used a long-short-term memory-based model for automatic bridge bidding. Although auction bridge games are traditionally played by two players, recently Sarkar et al.~\cite{ourpaper} recommended a novel gaming strategy for auction bridge that can accommodate three players at a time. This has considerable potential in expediting knockout tournaments of auction bridge, as well as adaptation in an online platform. 

After the bidding phase, gameplay proceeds similarly to traditional Auction Bridge; however, the strategies used are different, prioritizing a \textit{high-risk, high-reward} policy.

The order of calling in this game is as follows:
\[
\clubsuit \ (\text{Clubs}), \ \diamondsuit \ (\text{Diamonds}), \ \heartsuit \ (\text{Hearts}), \ \spadesuit \ (\text{Spades}), \ \text{No Trump},
\]
Bidding starts from 1 Clubs and ends with 7 No Trump.

Rotational bidding is allowed, similar to traditional bridge. However, in this game, three players bid among themselves, and the fourth hand automatically joins the winning bidder as their partner. In comparison, the remaining two players join the opposing team. In this manner, each game's bidding phase dynamically decides the partners and opponents.

All points and rules used in this game were collected from the Sarkar et al.~\cite{ourpaper}.

Sarkar et al.~\cite{ourpaper} discussed various types of game strategies such as \textit{defensive approach}, \textit{attacking approach}, \textit{defeat-seeking approach}, and \textit{bluff approach}. We calculated the probability of winning any hand with calls of 1, 2, and 3 no trump, considering only the presence of high face cards in the hand (specifically in the sequence: Ace, King, Queen, Jack, 10).

However, it was observed that having all high face cards is not always necessary to win a no-trump game. The computed probabilities to win 1, 2, and 3 no trump were $14 \times 10^{-5}$, $3.6 \times 10^{-5}$, and $0.5 \times 10^{-5}$, respectively, which are very low.

In contrast, if a player uses the \textit{low card first strategy} when not all high cards are present in the bidder's hand and uses the \textit{high card first strategy} when all high cards are present, as proposed in the previous paper, the probability of winning increases by a factor of one thousand.

In this paper, we discuss the algorithm for calling no trump. Unlike previous approaches that depend on exact cards, the proposed method relies on a set of defined rules. We establish these rules for no-trump calls and use strategies such as \textit{low card first} and \textit{high card first} based on the cards in hand, ensuring that the player can win the game almost certainly. It is important to note that although this algorithm identifies highly favorable bidding and winning situations as it is based upon the presence of cards in the hand of a bidder, the probability that it would allow a player to bid 1NT (and other NT calls) has a fairly low probability, which has been checked empirically via the simulations. The necessary files for the simulations are available at ~\cite{SarkarMajumdar2025AuctionBridgeAlgorithm}

Furthermore, complexity analysis is provided, demonstrating that dynamic programming is not required for this problem, and an enumerative approach is sufficient to determine the optimal call. Simulated data is presented to support and illustrate the improved results.

The rest of the article is arranged as follows. Section 2 discusses the no-trump strategy as a part of a three-player auction bridge game. Section 3 discusses the action of taking a risk by bidding winner with a partner's hand. Finally, the article concludes in Section 4 with a consolidated remark on the proposed gaming strategies for three-player auction bridge.

\section {No Trump Strategy}
In the Auction Bridge card game, the trump strategy plays a big role. As we discussed the point table and rules in the introduction section, if a player chooses No Trump instead of any fixed trump, then the player will earn more points. But, in three-player auction bridge, it's a very tough decision to decide on no trump instead of fixed trump because one player didn't call anything. If a certain color of cards is present in a hand, then the player should focus on fixed trump. If a player wants to call no trump, there are 2 cases:
\begin{itemize}
  \item The player's hand is good, but the player should depend on the partner's hand.
  \item The player is fully confident in his hand to win the game; there is no dependency on the partner's hand.
\end{itemize}

\subsection {Algorithm for no trump call}

A player completed 1, 2, or 3 no-trump calls without depending on the partner's hand. As per the previous paper, the probabilities of winning 1NT, 2NT, and 3NT are approximately $14 \times 10^{-5}$, $3.6 \times 10^{-5}$, and $0.5 \times 10^{-5}$, respectively. 

In this paper, we introduce an exact algorithm that helps a defensive player call no trump and win the game almost surely. Here $P$ stands for the number of tricks to be won according to the bid, and $K = 13 - P$, i.e the number of lost tricks the bidder can afford. Check~\ref{K-P table} for further clarification. In the simulation code present in~\cite{SarkarMajumdar2025AuctionBridgeAlgorithm}, the \texttt{call\textunderscore count} variable is the representative of the variable $P$.
\begin{algorithm}
\caption{No Trump algorithm}
\label{alg:notrump}
\begin{algorithmic}[1]
\State \textbf{Input:} Standard 52-card deck
\State \textbf{Output:} Decision to proceed based on conditions

\State \textbf{Initialize:} suits $\gets$ [Clubs, Diamonds, Hearts, Spades], ranks $\gets$ [Ace, King, Queen, Jack, 10,\dots, 2]
\State $deck \gets$ all (suit, rank) pairs
\State $hand \gets$ randomly select 13 unique cards from $deck$

\State Initialize the dataframe $df$ with suits as rows, and ranks as columns, with all entries = 0.
\For{each (suit, rank) in $hand$}
    \State $df[suit, rank] \gets 1$
\EndFor

\For{each row in $df$}
    \State $count\_ones \gets$ number of 1s in the row
    \If{$count\_ones == 0$}
        \State $zeros\_before\_first\_one \gets 13$
    \Else
        \State $zeros\_before\_first\_one \gets$ index of the first 1
    \EndIf
    \State Store $zeros\_before\_first\_one$ and $count\_ones$ in $df$
\EndFor

\State $total\_zeros\_before\_first\_1 \gets$ sum of all $zeros\_before\_first\_one$ across suits

\If{$total\_zeros\_before\_first\_1 \leq K$}
    \State \Comment Condition A passed
    \State $condition\_B\_passed \gets$ True if for all rows, $count\_ones > zeros\_before\_first\_one$
    
    \If{$condition\_B\_passed$}
        \State Initialize $count\_of\_one \gets 4$
        \State Initialize $remaining\_zeros \gets K - total\_zeros\_before\_first\_1$
        
        \For{each suit in suits}
            \State $indices\_of\_ones \gets$ indices of 1s in row
            \For{$i=0$ to $len(indices\_of\_ones)-2$}
                \State $zeros\_between \gets indices\_of\_ones[i+1] - indices\_of\_ones[i] - 1$
                
                \If{$zeros\_between \leq remaining\_zeros$}
                    \State $count\_of\_one \gets count\_of\_one + 1$
                    \State $remaining\_zeros \gets remaining\_zeros - zeros\_between$
                \EndIf
                
                \If{$remaining\_zeros < 0$}
                    \State \textbf{break}
                \EndIf
            \EndFor
            \If{$remaining\_zeros < 0$}
                \State \textbf{break}
            \EndIf
        \EndFor

        \If{$count\_of\_one \geq P$}
            \State \Return Yes, final condition satisfied. Go ahead.
        \Else
            \State \Return No, final condition failed.
        \EndIf
    \Else
        \State \Return Condition B failed. Do not proceed.
    \EndIf
\Else
    \State \Return Condition A failed. Do not proceed.
\EndIf

\end{algorithmic}
\end{algorithm}

\begin{table}[h!]
\centering
\caption{Number of No Trump Calls Depends on K and P}
\begin{tabular}{|*{3}{D|}}
\hline
NT & K & P \\
\hline
1 & 6 & 7 \\
2 & 5 & 8 \\
3 & 4 & 9 \\
\hline
\end{tabular}
\label{K-P table}
\end{table}

This algorithm can be divided into three parts.
Initially, create a $13\times4$ $(m\times n)$ dataframe and put 1 if the card exists in the player's hand, otherwise 0. So, a total of 13 ones and 39 zeros exist in that dataframe. 
  \begin{center}
      \begin{tabular}{|l|*{13}{C|}}
\hline
 & A & K & Q & J & 10 & 9 & 8 & 7 & 6 & 5 & 4 & 3 & 2 \\
\hline
Clubs &  &  &  &  &  &  &  &  &  &  &  &  &  \\
\hline
Diamonds &  &  &  &  &  &  &  &  &  &  &  &  &  \\
\hline
Hearts &  &  &  &  &  &  &  &  &  &  &  &  &  \\
\hline
Spades &  &  &  &  &  &  &  &  &  &  &  &  &  \\
\hline
\end{tabular}
  \end{center}

\begin{itemize}
  \item \textbf{Step 1:} Check for each of the 4 rows, from left to right, how many 0s exist before the first 1. If the first one is 1, then the number of zeros before the first 1 is 0. If no one is present, then the output should be 13. Furthermore, add how many 1s are present in each of the 4 cases.
  \item \textbf{Step 2:}  Calculate the sum of zeros before the first 1; if it is less than or equal to 6, then we go ahead to the next step.
  \item\textbf{Step 3:} Check the total number of 1s in each row is greater than or equal to zero before the first 1. If this condition is satisfied for every 4 rows, then go ahead; otherwise, do not.
  \item\textbf{Step 4:} If both the previous conditions are satisfied, then we go ahead to the next step in the nested previous condition. Take a new variable (count of one) initialized at 4, and the sum of zeros is already known. Initialize a new variable (remaining zeros), which is equal to ($K$-sum of zeros). Now from the leftmost 1 of the first row, move to the next 1 and count the zeros between. If the count is greater than the remaining zeros, don't do anything; just go ahead. If the count is less than or equal to remaining zeros, add count 1 with a count of one, and subtract that many zeros from remaining zeros while the \textit{loop condition} remaining zeros is greater than or equal to 0. Now traverse all 4 rows. And finally, check if the count of one is greater than or equal to $P$ or not. If yes, then print \texttt{YES}, otherwise print \texttt{NO}.
\end{itemize}

\begin{table}[h!]
\centering
\caption{Complexity Analysis of No Trump Algorithm}
\begin{tabular}{|c|l|c|}
\hline
\textbf{Step} & \textbf{Description} & \textbf{Time Complexity} \\
\hline
1 & Deck creation (m suits × n ranks) & $O(mn)$ \\
2 & Random sampling of 13 cards & $O(1)$ \\
3 & data frame initialization & $O(mn)$ \\
4 & Updating the dataframe with hand cards & $O(1)$ \\
5 & Analyzing each row & $O(mn)$ \\
6 & Summing zeros\_before\_first\_1 & $O(n)$ \\
7 & Condition B check & $O(n)$ \\
8 & Final nested loops (suits × indices\_of\_ones) & $O(nm)$ \\
\hline
 & \textbf{Total} & $O(mn)$ \\
\hline
\end{tabular}
\end{table}

\begin{table}[h!]
\centering
\caption{Simulation Results for 1NT, 2NT, and 3NT (100,000 Simulations Each)}
\begin{tabular}{|c|c|c||c|c|c||c|c|c|}
\hline
\multicolumn{3}{|c||}{\textbf{Simulation for 1NT - 100,000}} & \multicolumn{3}{c||}{\textbf{Simulation for 2NT - 100,000}} & \multicolumn{3}{c|}{\textbf{Simulation for 3NT - 100,000}} \\
\hline
Sim no & Yes & No & Sim no & Yes & No & Sim no & Yes & No \\
\hline
1 & 5955 & 94045 & 1 & 3228 & 96772 & 1 & 875 & 99125 \\
2 & 5992 & 94008 & 2 & 3174 & 96826 & 2 & 844 & 99156 \\
3 & 5849 & 94151 & 3 & 3186 & 96814 & 3 & 850 & 99150 \\
\hline
\end{tabular}
\end{table}

\begin{figure}[h!]
    \centering
    \includegraphics[width=0.7\textwidth]{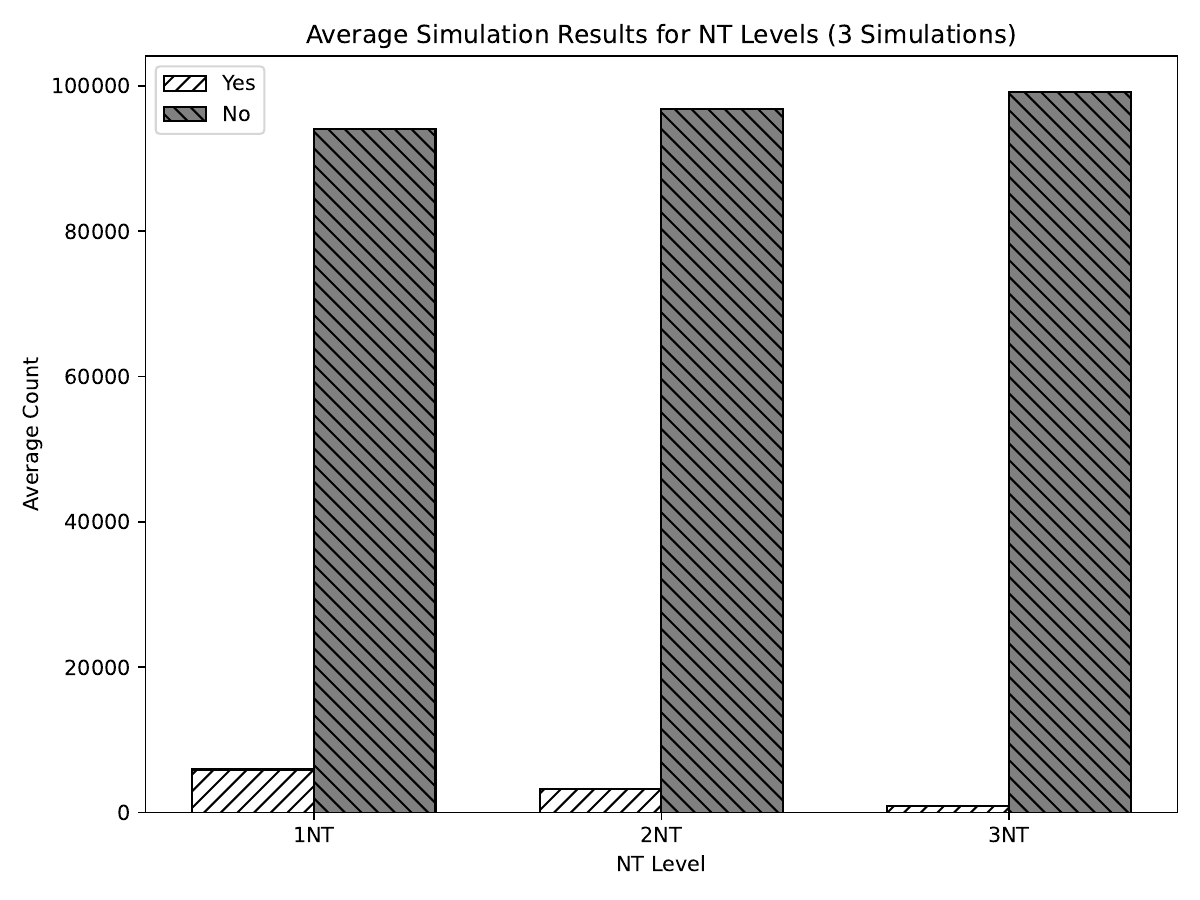}
    \caption{Average simulation results for 1NT, 2NT, and 3NT over 3 simulations each.}
    \label{fig:nt_simulation}
\end{figure}

The probabilities of winning 1 No Trump, 2 No Trump, and 3 No Trump are approximately 0.06, 0.03, and 0.0085, respectively.
\newpage
The algorithm essentially tracks the total number of missing high cards by the `total number of zeroes before the first one' and also calculates the number of tricks that can be obtained by the bidder based on the cards available in hand. Thus, it determines the \emph{success} of a \textbf{NO Trump} call.

A 1NT opening is relatively common because it is used to describe a balanced hand within a narrow high-card-point range (typically 15-17 HCP in modern systems). In contrast, a 3NT opening is a preemptive or highly specialized call, usually requiring a much stronger hand with sufficient stoppers in all suits and a high likelihood of making nine tricks without exploring a major-suit or slam contract. Such hands occur far less frequently than the balanced hands suitable for a 1NT opening. Consequently, 3NT openings are significantly rarer than 1NT openings. This is also evident from the results of the simulations performed.

\section{Simulation Methodology}
The proposed no-trump bidding algorithm was implemented in \texttt{Python} using the  \texttt{pandas} and \texttt{random} libraries. To evaluate its performance, a Monte Carlo simulation framework was employed. For each simulation, a standard 52-card deck was generated, and a 13-card bridge hand was sampled uniformly at random without replacement to represent the hand of the bidder. The rest of the code is based upon the algorithm mentioned before.

\section{Bidding Winner with Partner's Hand}
After bidding, the winning bidder automatically chooses the 4th hand as a partner.
A defensive strategist winning bidder doesn't depend on the partner's hand to win the game. But most of the time, bidders take some risk and depend on the partner's hand. The winning bidder knows 26 cards after bidding, and winning the opportunity is highly increased after starting the game. 

\begin{table}[h!]
\centering
\caption{No Trump Results: Single Hands vs. Partners}
\begin{tabular}{|c|r|r|r|}
\hline
\textbf{No Trump} & \textbf{Single Hand} & \textbf{With Partner} & \textbf{Percentage} \\
\hline
1 & 6005  & 79991 & 0.075 \\
2 & 3259  & 73633 & 0.044 \\
3 &  814  & 63276 & 0.013 \\
\hline
\end{tabular}
\label{tab:no_trump_results}
\end{table}

\begin{figure}[htbp]
    \centering
    \includegraphics[width=0.55\textwidth]{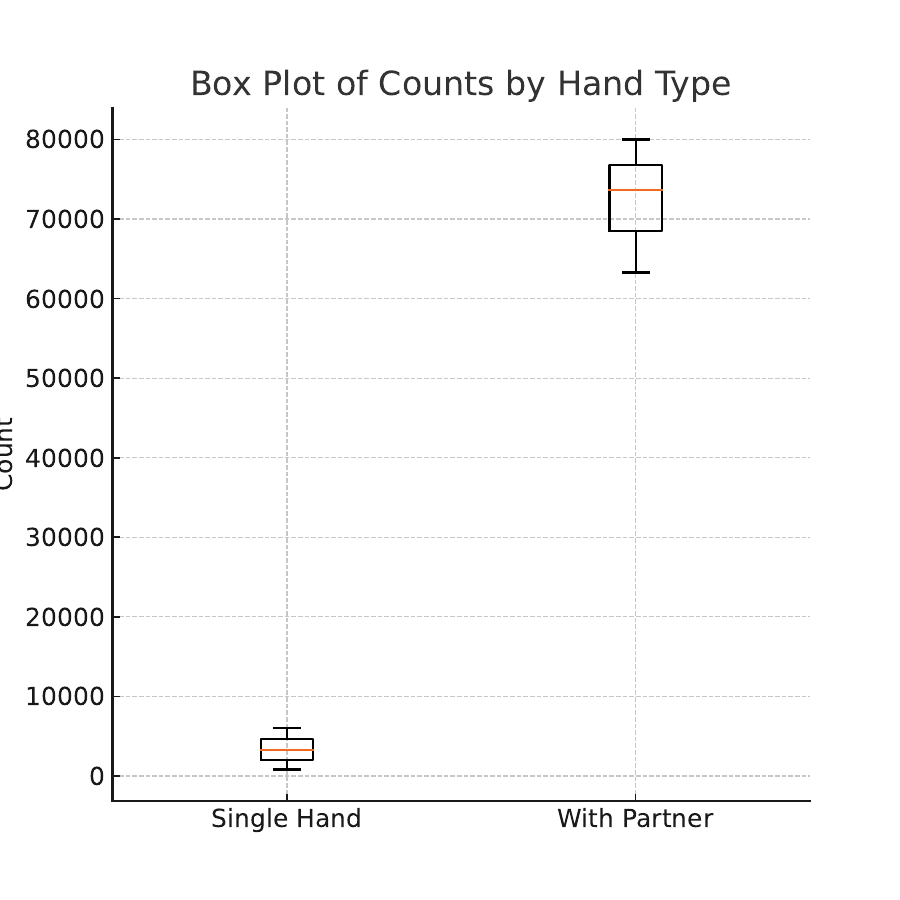}
    \caption{Box Plot of Counts by Hand Type (Single Hand vs. With Partner)}
    \label{fig:box_plot_hand_type}
\end{figure}

\section{Conclusion}
The three-player auction bridge variant introduces dynamic partnerships and a unique bidding structure, significantly enriching strategic complexity. Through rigorous algorithmic development and simulation, this study demonstrates that a deterministic, rule-based approach to no-trump calls can drastically improve success rates—outperforming earlier probabilistic heuristics by several orders of magnitude. Although the defensive strategy guarantees safe play, the simulations reveal that attacking and bluff strategies, particularly when optimized through algorithmic insight, often yield superior results. Our findings underscore that in this dynamic environment, calculated risks—especially when supported by intelligent evaluation of hand strength and partner potential—can lead to consistently better outcomes.Moreover, there exists another interesting strategy, a \emph{defeat-seeking} strategy, and that has been mentioned in \cite{ourpaper}. This particular algorithm works better than any machine learning approach used for \emph{no trump calls} with significantly less time complexity. This study lays the groundwork for future enhancements in algorithmic bidding and adaptive game strategies in multiplayer card games. Moreover, it is interesting to note that the algorithm calls 1 NT in four out of five games successfully if \emph{full information} about both of the hands of a team is available. Later, our plan is to modify this so that calls can be given even with the partial information obtained from the calls given by the partner.
\bibliographystyle{splncs04}
\bibliography{ref}
\end{document}